\documentclass[10pt]{aastex62}

\usepackage{graphicx}
\usepackage{aas_macros}

\usepackage{float}

\usepackage{algorithm2e}
\usepackage{natbib}

\begin{document}
 \shorttitle{EM counterparts of BBH coalescences}
  \shortauthors{Noysena et al.}
\title{Limits on the Electro-Magnetic Counterpart of Binary Black Hole Coalescence at Visible Wavelengths}

\author[0000-0001-9109-8311]{Kanthanakorn Noysena}
\affiliation{Institut de Recherche en Astrophysique et Plan\'etologie (IRAP), Universit\'e de Toulouse, CNRS, UPS, CNES, 14 avenue Edouard Belin, Toulouse, 31400, France}
\affiliation{CNRS, UCA, OCA, ARTEMIS UMR 7250, boulevard de l'Observatoire, F 06304 Nice Cedex 4, France}
\author{Alain Klotz}
\affiliation{Institut de Recherche en Astrophysique et Plan\'etologie (IRAP), Universit\'e de Toulouse, CNRS, UPS, CNES, 14 avenue Edouard Belin, Toulouse, 31400, France}
\author[0000-0001-9157-4349]{Michel Bo\"er}
\affiliation{CNRS, UCA, OCA, ARTEMIS UMR 7250, boulevard de l'Observatoire, F 06304 Nice Cedex 4, France}
\author{Romain Laugier}
\affiliation{CNRS, UCA, OCA, ARTEMIS UMR 7250, boulevard de l'Observatoire, F 06304 Nice Cedex 4, France}
\author{Siramas Komonjinda}
\affiliation{Research Center of Physics and Astronomy, Faculty of Science, Chiang Mai University, Chiang Mai, 52000, Thailand}
\author{Damien Turpin}
\affiliation{National Astronomical Observatories, Chinese Academy of Sciences, Beijing 100012, China}
\affiliation{Institut de Recherche en Astrophysique et Plan\'etologie (IRAP), Universit\'e de Toulouse, CNRS, UPS, CNES, 14 avenue Edouard Belin, Toulouse, 31400, France}
\collaboration{(The TAROT Collaboration)}
\correspondingauthor{Kanthanakorn Noysena}
\email{Kanthanakorn.Noysena@irap.omp.eu}

\received{March 28, 2019}
\revised{}
\accepted{}

\begin{abstract}
We used the TAROT network of telescopes to  search for the electromagnetic counterparts of GW150914, GW170104 and GW170814, which were reported to originate from binary black hole merger events by the LIGO and Virgo collaborations. Our goal is to constrain the emission from a binary black hole coalescence at visible wavelengths. We developed a simple and effective algorithm to detect new sources by matching the image data with the Gaia catalog data release 1. Machine learning was used and an algorithm was designed to locate  unknown sources in a large field of view image. The angular distance between objects in the image and in the catalog was used to find new sources; we then process the candidates to validate them as possible new unknown celestial objects. Though several possible candidates were detected in the three gravitational wave source error boxes studied, none of them were confirmed as a viable counterpart.
The algorithm was effective for the identification of unknown candidates in a very large field and provided candidates for GW150914, GW170104 and GW170814. The entire 90\% GW170814 error box was surveyed extensively within 0.6 days after the GW emission resulting in an absolute limiting R magnitude of -23.8. This strong limit excludes to a great extent a possible emission of a gamma-ray burst with an optical counterpart associated with GW170814.
\end{abstract}
\keywords{gravitational waves, stars: black holes, gamma-ray burst: general}


\section{INTRODUCTION}\label{sec:intro}
	The Laser Interferometer Gravitational-wave Observatory (LIGO) detected gravitational waves (GW) for the first time on September 14, 2015~\citep{PhysRevLett.116.061102}. Called GW150914 the analysis of the source signal was compatible with the merging of a binary system of black holes (BBH) of masses $M_{1} = 36^{+5}_{-4}M_{\odot}$ and $M_{2} = 29^{+4}_{-4}M_{\odot}$. The localization of GW150914 has been sent to 25 teams operating ground and space based telescopes. No electromagnetic (EM) counterpart was found nor on ground~\citep{abbott:in2p3-01283976}  nor in space, excepted, possibly, in one case~\citep{Connaughton:2016umz}. A binary black hole is usually believed to be almost free of surrounding matter, preventing detectable EM emission~\citep{Diaz:2016yly}. 
	EM waves are closely coupled to matter thus they provide significant information about the environment of the progenitor system, e.g. an accretion disk, and the physical process at work during and after the coalescence~\citep{Burns:2018pcl, Zhang:2016kyq}. BBHs can be produced either by the evolution of massive stars~\citep{Belczynski:2016obo} in a binary system or by dynamical formation in globular clusters~\citep{2016ApJ...824L...8R}. The relative importance of both routes remains debated. It has been proposed that if at least one of two black holes has enough quantity of charge retained by a rotating magnetosphere then a rapid merger's evolution could drives a Poynting flux to power a SGRB or an optical transient that would be detectable~\citep{Zhang:2016rli}. Another process invoked by \citet{Stone:2016wzz} is based on BBH formation in the self-gravitating disks of active galactic nuclei (AGN). The authors proposed that these conditions can favor an EM counterpart due to super-Eddington accretion onto the black hole following the merger.
	The detection of an EM counterpart of a BBH event, or the limits we can derive from the absence of such detection, would contribute to the understanding of  BBH merger processes and it would provide clues on its localization and history, e.g. evolution from a field binary system or dynamic evolution in a globular cluster.
	
	The LIGO signal analysis showed that an energy of $\Delta E = 3.0^{+0.5}_{-0.4}M_{\odot}$ was released during the coalescence of GW150914. 
	Even if a small fraction of $\Delta E$ of this energy was to be emitted into EM radiation, it could be eventually detected by ground facilities. The corresponding luminosity L is given by Equation \ref{eq:1}:
    \begin{equation}
	    \label{eq:1}
	{%
		L = \frac{\alpha \Delta E}{\Delta t}
	}
	\end{equation}
	where $\alpha$ is the fraction of the $\Delta E$ energy converted into EM radiation $(0 < \alpha \ll 1)$ and $\Delta t$ is the emission duration.
	A rough estimation of the optical luminosity is made considering a constant solar black body emission from the date of the trigger until the date of the observation. Knowing the Sun luminosity $L_{\odot}$ = $4\times10^{26} \rm W$ and distance $D_{\odot}$ = $1.5\times10^{11} \rm m$, we can estimate the optical magnitude from Equation~\ref{eq:2}:	
	
	\begin{equation}
		\label{eq:2}
	{%
		m_{candidate} = -16.12 - 2.5log\frac{L}{D_{GW}^{2}} 
	}
	\end{equation}
	
	where $D_{\textsc{gw}}$ is the luminosity distance of the GW source. \textbf{If we suppose that} the optical counterpart has a magnitude of 14, 20 minute after the beginning of the event, then $\alpha=3\times10^{-7}$; If the emission lasts 2 days, when observations are proceeding, then the parameter $\alpha$ will become $5\times10^{-5}$. Small aperture telescopes are adequate for the detection and follow-up of a possible optical transient event because they can probe values of $\alpha$ $\ll$1. 
	
	During the LIGO and Virgo O1 and O2 runs~\citep{PhysRevLett.119.161101, 2018arXiv181112907T}, several BBH events were detected with their coordinates sent to the observer community: GW150914~\citep{PhysRevLett.116.061102}, LVT151012~\citep{PhysRevX.6.041015}, GW151226~\citep{PhysRevLett.116.241103}, GW170104~\citep{PhysRevLett.118.221101}, GW170608~\citep{ffb939f2d7de4d9290d876983c0a15f8}, and GW170814~\citep{PhysRevLett.119.141101}. The difficulty to find the optical transient of BBH mergers comes from the very large area covered by the GW error boxes generated by LIGO, which covers typically 1000 $deg^{2}$ albeit these areas can be efficiently scanned  with very wide field of view (FoV) telescopes. The beginning of the operations of Virgo~\citep{PhysRevLett.119.141101} in August 2017 resulted in a considerable reduction in size, about 60 $deg^{2}$ for GW170814 for the 90\% error region. Searches for optical counterparts of GW events started with the first GW detection thanks to the GCN support, which disseminated quickly probability sky-maps. Searches for a possible optical transient associated with GW150914 were performed by∼\cite{Brocato:2017bdr} with the 2.6m VST telescope, ~\cite{Soares-Santos:2016qeb} with the Dark Energy Camera (DECam) and ~\cite{Kasliwal:2016uhu} with the intermediate Palomar Transient factory (iPTF). All these telescopes covered less than 50\% of the error region, reaching a limiting magnitude of $\sim21.0$ to 22.7; they started their observations $\sim4$ days after the event. \cite{Stalder:2017qic} scanned the error box of GW170114 with the Asteroid Terrestrial-impact Last Alert System (ATLAS): the covered 43\% of the error box, 23.1 hours after the event with a limiting magnitude of $\sim21.5$ in i-band. \cite{Turpin:2019bjj} observed 62\% of the GW170104 error box with the mini-GWAC telescopes reaching a limiting magnitude of 16 2.3 hours after the event. \cite{Doctor:2018ray} observed the GW170814 skymap with DECam covering 86\% of the error box, 1.0 day after the event with the limiting magnitude of $\sim23$ in i-band. No optical counterpart of these GW events has been identified.
	
	TAROT (T\'{e}lescope \`{a} Action Rapide pour les Objets Transitoires -- Rapid Action Telescope for Transient Objects) is an automated robotic telescope network~\citep{1538-3873-120-874-1298,boer:hal-01511688} that has a very large FoV combined with a fast response (less than 10s) to alert notices. These instruments have been used since 1998 to perform early optical observations of Gamma-Ray Bursts~\citep{1538-3881-137-5-4100}. The TAROT network participated to the first prompt search for GW transient EM counterparts organized in 2010 by the LIGO and Virgo teams~\citep{2012A&A...539A.124L}. Three instruments were used during the LIGO-Virgo (LV) runs O1 and O2 between 2015 and 2017: TAROT Calern (TCA, long. = 6.92353 E lat. = 43.75203 alt. = 1320m), TAROT La Silla (TCH, long. = 70.73260 W lat. = -29.25992 alt. = 2398m), TAROT Reunion (TRE, long. = 55.41022 E lat. = -21.19882 alt. = 991m). TCA and TCH have an aperture of 250mm and a FoV $1.8\times1.8$ $\rm deg^{2}$. Both telescopes are equiped with  ANDOR Ikon L936 back illuminated CCD cameras. TRE is a commercial instrument with an aperture of 180mm equiped with a the FLI Proline KAF-16803 CCD camera, resulting in a FoV of $4.2\times4.2$ $deg^{2}$. TCA and TCH have a limiting R magnitude of 18.2 at $5\sigma$ for 1 minute  unfiltered exposure~\citep{2009AIPC.1133..175G}; the limiting R magnitude of TRE is 17.	Thanks to the large FoV of the TAROT instruments, and their different locations over the Earth, it is possible to cover a large part of the GW error box in a relatively short time. As an example, TRE can observe 50 $deg^{2}$ within 10 minutes. However that means a huge amount of data, which should be processed immediately after acquisition. We elaborated a simple, fast and reliable algorithm to search and identify possible transient counterpart candidates.

	The luminosity distance of the  known BBH GW events derived from the gravitational wave analysis range from $320_{-110}^{+120}~\rm Mpc$ to $2750_{-1320}^{+1350}~\rm Mpc$ with a median of $\approx 900$~Mpc~\citep{2018arXiv181112907T}. Considering a conservative limiting magnitude of 17 for a small aperture telescope, the corresponding absolute limiting magnitude is -22.8. At this level, only the afterglow of GRBs can be detected: supernovae, flares, kilonovae are too faint. As a consequence this study is limited to the search for an association between GW events and GRB optical counterparts, i.e. optical transients.
	
	In this paper we present the search for transient counterpart candidates associated with GW150914, GW170104 and GW170814. We designed a simple algorithm that uses machine learning and available public catalogs. In the next section we describe the method that we used to process TAROT images. The third and fourth sections summarize the data observed by TAROT in the error box of the three above mentioned GW events; we perform the search for transient source candidates and we evaluate  the performance of our extraction procedure. In the fifth section we discuss the results and present a strong limit on the possible optical transient counterpart of GW170814, before our concluding remarks in the last section.

\section{THE METHOD}\label{sec:method}
After the analysis of the images, we compare the sources that are found within the data with the Gaia catalog data release 1 \citep{2016A&A...595A...4L}. To speed up the process we implemented the catalog in a local machine. The traditional  approach is to cross-match the presence of the possible objects in a catalog within a given error region that depends on the instrument, acquisition conditions, and of a catalog~\citep{1992A&A...258..217E}. Another approach is to cluster objects hierarchically in a dendrogram before comparison: the hierarchical clustering methods that are either of the "agglomerate type", such as single-linkage~\citep{2012ApJ...761..188B}, or of the "divisive type" such as k dimensional tree (K-d tree) by~\citet{Bentley:1975:MBS:361002.361007}, are often used in computational data to illustrate the clustering of samples. In this work we used the K-d tree algorithm as it uses less computing time (we need to send our results timely for the follow-up of our candidates by other, larger, telescopes), and its lower complexity requires less memory. This algorithm performs also a quick search for the nearest neighbors for any data coordinate. The algorithm is available in scipy.spatial.KDTree~\citep{1999cs........1013M} and it has been implemented in \textit{math\textunderscore coordinates\textunderscore sky} function of the Astropy package by~\citet{2013A&A...558A..33A}.

\begin{figure}[h]
        \begin{center}
        \includegraphics[width=0.55\linewidth]{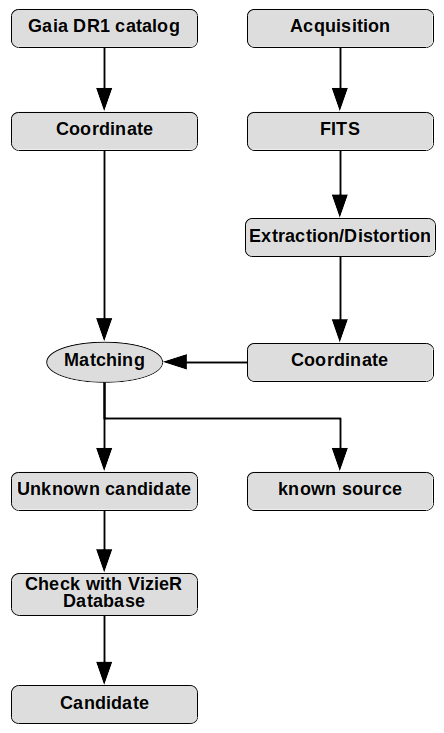}		    
	    \end{center}
	    \caption{Schematics of the processing.}
	    \label{fig:Fig1}
    \end{figure}

Our procedure is summarized in Figure~\ref{fig:Fig1}; it is based on the following 5 steps:
\begin{enumerate}
    \item[1:] After acquisition the image is calibrated for non uniformity and distortions using the Astrometry.net package~\citep{2010AJ....139.1782L}; we use index files from the Tycho-2 and 2MASS catalogs; this algorithm compares the shape of sets of four stars with the shape of reference stars in the index files and it computes the World Coordinate System (WCS) coefficients with the Simple Imaging Polynomial (SIP) convention.
    \item[2:] The SExtractor package provided by~\citet{1996A&AS..117..393B} has been used to extract sources in TAROT images; this algorithm works by determining the background and whether pixels belong to background or objects before it splits up objects from background; we specifically require pixel coordinates from the image.
    \item[3:] The pixel coordinates of the sources are converted to equatorial coordinates using the tools available in the Astropy.wcs package from~\citet{2013A&A...558A..33A} and the WCS and SIP coefficients obtained from the second step.
    \item[4:] We use the K-d tree algorithm to match each source with the Gaia DR1. The catalog coordinates in the FoV are used as training data by recursively partitioning the data set: we look for source coordinates in the same FoV through the data set at the nearest neighbor point. The median of the angular separation and the vector of the direction of a match are computed and applied to shift the image and reiterating until the median of the angular separations is minimized. If a source has an angular separation higher than the median then it is considered as a mismatch and classified as an unidentified source.
    \item[5:] We then flag this Gaia unidentified source as a possible candidate if it is not present in the USNO-B1.0 catalog, after retrieval of the updated position from the online VizieR database~\citep{2000A&AS..143...23O}. We apply eventually the photometric criterion for decay parameter consideration.
\end{enumerate}

\begin{figure}[h]
        \begin{center}
	    \includegraphics[width=1.0\linewidth]{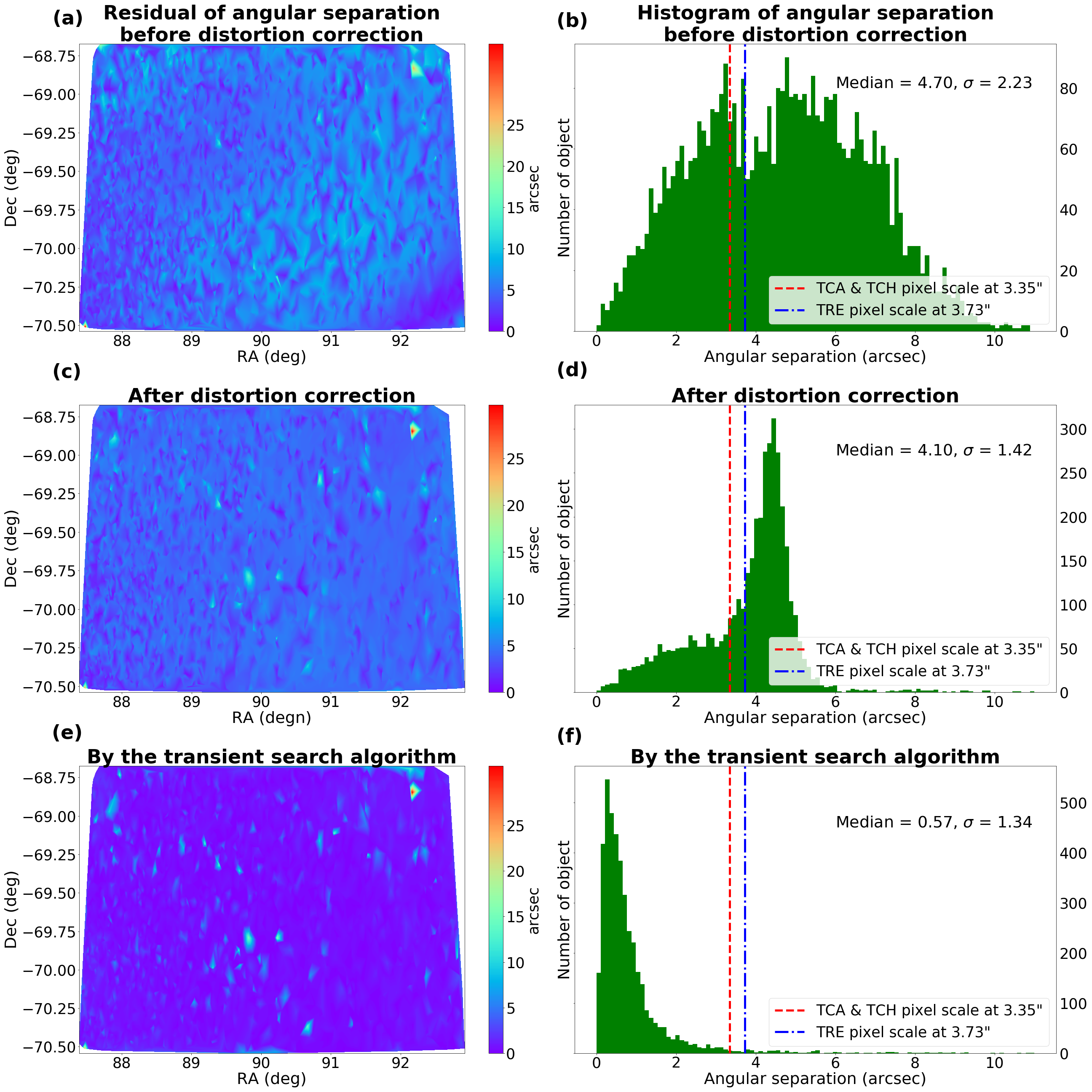}
        \end{center}
        \caption{Distortions in a typical TAROT image obtained from the comparison with the Gaia DR1 catalog. (a) The distortion after the first direct Gnomonic projection and (b) the histogram of the angular separation; (c) the distortion after the correction computed using the SIP5 algorithm and (d) the corresponding histogram; (e) After correction by the transient search algorithm, (f) the histogram shows that we reached a sub-pixel accuracy.  The red dashed line and blue dash-dotted line is the pixel size of TCA and TCH, the blue dash-dotted line is the pixel size of TRE. On the left the color code is from red (high separation) to dark blue (exact match).}
        \label{fig:Fig2}
    \end{figure}

As TAROT has a large FoV, images are subject to optical distortions which shows in the residues of angular separations between WCS and the astrometric catalog coordinates in Figure~\ref{fig:Fig2}a and~\ref{fig:Fig2}b. The color code shows the angular separation as from dark blue (exact match) to red. However a distortion can be corrected by applying the Simple Imaging Polynomial convention~\citep{2004ASPC..314..551C} at order $5^{th}$~(SIP5)~\citep{2005ASPC..347..491S} in FITS WCS and the result is displayed in Figure~\ref{fig:Fig2}c, showing a much more homogeneous residue map than that of Figure~\ref{fig:Fig2}a nevertheless the median residue is still at $4.10''$ in Figure~\ref{fig:Fig2}b, higher than the pixel scale of TAROT; $3.35''$/pixel for TCA and TCH and $3.73''$/pixel for TRE~\citep{boer:hal-01511688}. In order to get the smallest angular separations and to reach the ultimate, sub-pixel, precision for TAROT, we iterated the matching procedure. As a result the median separation gets much smaller, as shown in~\ref{fig:Fig2}e and~\ref{fig:Fig2}f. possible candidates are located by the search algorithm and then checked against the USNO-B1.0 catalog~\citep{2003AJ....125..984M} within a search radius of $10~arcsec$. If we consider that the flux is proportional to t$^{-\alpha_{opt}}$, as it is the case for decays of GRB afterglows, we can compute the decay parameter $\alpha_{opt}$ using Equation~\ref{eq:3} for the candidates that have measured magnitudes $m_{1}$ and $m_{2}$ at two epochs $t_{1}$ and $t_{2}$:

	\begin{equation}
		\label{eq:3}
	{%
	   \alpha_{opt} = \frac{m_{1} - m_{2}}{2.5\times\log\big(\frac{t_{1}-t_{trig}}{t_{2}-t_{trig}}\big)}
	}
	\end{equation}
Where $t_{trig}$ is the trigger time of GW event. We considered that GRBs have decays $0.5<\alpha_{opt}<2.5$. All candidates that do not satisfy this criterion are rejected.

\section{OBSERVATIONS}\label{sec:observation}
    Three GW events related to BBH mergers were observed by the TAROT network: GW150914, GW170104 and GW170814. As the FoV of the instruments is small compared to the size of the error box, we divided it in tiles covering the area of probability levels above 10\%. The tiles were chosen in order to observe them at least three times during the night, to avoid fake detection and to get comparison images (Table \ref{table:tab1}).
        
\begin{figure}
        \begin{center}
        \includegraphics[width=1\linewidth]{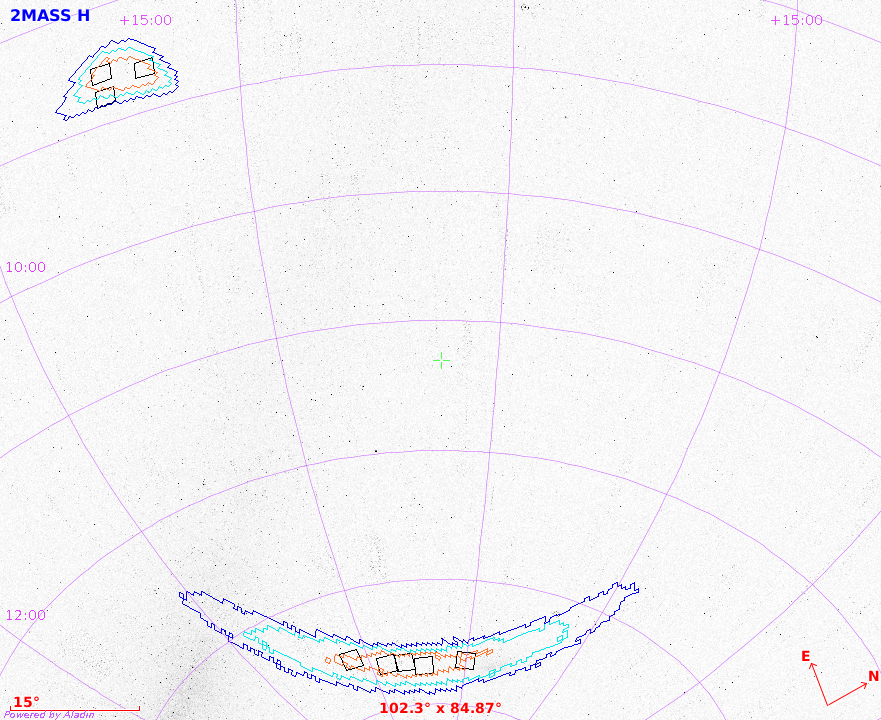}
        \caption{The tiles observed by TAROT are displayed over the contours of the initial distributed localization of GW150914. Each square represents a FoV of 3.24 $\rm deg^{2}$. The lines represent the enclosed 90\%, 70\% and 30\% probability contour levels. TAROT observed $\sim$8\% of the GW initial error box.}
	    \label{fig:gw150914_skymap}
        \end{center}
    \end{figure}

    The GW150914 error box was observed by TCA and TCH from three days after the event, until September 30, 2014. A total of 8 tiles were observed as shown in Figure \ref{fig:gw150914_skymap}, resulting in 400 frames.

\begin{figure}
	    \begin{center}
	    \includegraphics[width=1\linewidth]{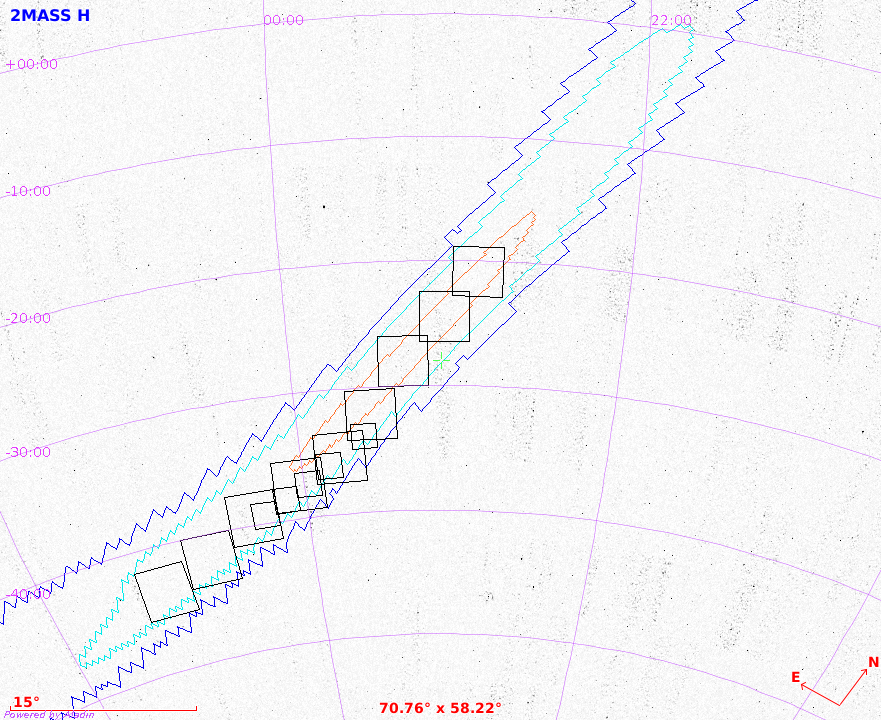}
	    \caption{The tiles observed by TCA and TCH (3.24 $\rm deg^{2}$), and TRE (17.98 $\rm deg^{2}$) are displayed over the contours of the localization of GW170104. The lines represent the enclosed 90\%, 70\% and 30\% probability contour levels of the initial BAYESTAR localization. Two third of the 70\% error box were observed by TAROT.} 
	    \label{fig:gw170104_skymap}
	    \end{center}
    \end{figure}

	The error box of GW170104 was observed by TCA, TCH and TRE within 30 hours after the GW detection until January 10. 18 tiles were observed repeatedly as shown in Figure \ref{fig:gw170104_skymap}. A total of 337 images were produced.

\begin{figure}
	    \begin{center}
	    \includegraphics[width=1\linewidth]{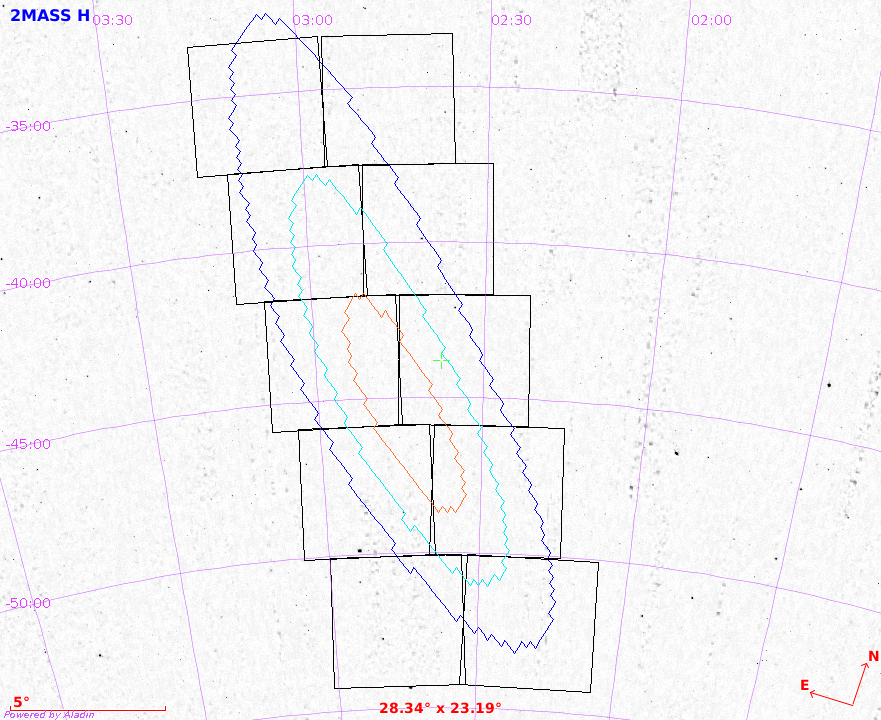}
	    \caption{The tiles observed by TRE (17.98 $\rm deg^{2}$) are displayed over the contours of the localization of GW170814. The lines represent the enclosed 90\%, 70\% and 30\% probability contour levels of the initial BAYESTAR localization. Almost the entire 90\% error box was observed by TRE.} 
	    \label{fig:gw170814_skymap}
	    \end{center}
    \end{figure}

	The GW170814 error box was observed by TCA, and TRE within 10 hours after GW detection until August 18. 13 pointing were repeated (Figure \ref{fig:gw170814_skymap}), resulting in a total of 333 images. 
	
	The exposure time for each observation was 120 seconds and no filter was used. The limiting magnitude was measured to be at least $\sim18$. We applied the procedure described above to all images.
	
	\begin{table*}[h]
	    \centering
	    \smaller
	    \caption{The log of the observations of the tiles derived from the GW error boxes by the TAROT network for GW150914, GW170104 and GW170814. The first column is the ID number of the tile; column 2 the telescope name, and column 3 the corresponding FoV; Columns 4 and 5 are the coordinates of the centre of the tile; column 6 is the time of the first acquisition with respect of the GW event time; column 7 is the limiting magnitude; column 8 is the conversion efficiency from GW to EM, as defined by the $\alpha$ factor in Equation~\ref{eq:1}.}
	    \begin{tabular}{@{}c cc cc cc c@{}}
       	\hline
       		\multicolumn{8}{c}{GW150914}\\
       	\hline
       		1&2&3&4&5&6&7&8 \\
       		Tile \# & Instrument & FoV & RA & Dec & $\Delta t$ & Limiting& Conversion \\
       		        &            &     &    &     &        &magnitude&efficiency\\
	       	& & ($\rm deg^{2}$) & (J2000) & J2000) & (day) & (R) & ($\times 10^{-6}$)\\
       	\hline
       		1 & TCH & 3.24 & 090.20 & -69.60 & 2.798 & 17.16 & 3.75 \\
       		2 & TCH & 3.24 & 103.50 & -70.30 & 2.805 & 16.10 & 9.99 \\
       		3 & TCH & 3.24 & 109.50 & -69.70 & 2.812 & 19.00 & 0.69 \\
       		4 & TCH & 3.24 & 115.00 & -69.60 & 2.854 & 18.62 & 0.99 \\
       		5 & TCH & 3.24 & 125.00 & -68.00 & 2.937 & 18.60 & 1.04 \\
       		6 & TCA & 3.24 & 130.10 & +05.30 & 3.736 & 17.01 & 5.72 \\
       		7 & TCA & 3.24 & 134.50 & +03.30 & 3.743 & 17.09 & 5.34 \\
       		8 & TCA & 3.24 & 134.70 & +06.30 & 3.749 & 17.13 & 5.15 \\
		\hline
		    \multicolumn{8}{c}{GW170104}\\
		\hline
		    1 & TRE & 17.98 & 343.07 & -20.85 & 1.255 & 17.33 & 11.03 \\
		    2 & TRE & 17.98 & 000.58 & -37.34 & 1.260 & 16.88 & 15.20 \\
		    3 & TRE & 17.98 & 011.15 & -42.02 & 1.265 & 18.24 & 4.34 \\
		    4 & TRE & 17.98 & 352.89 & -32.19 & 1.273 & 16.75 & 17.29 \\
		    5 & TRE & 17.98 & 345.98 & -24.51 & 1.277 & 16.82 & 16.22 \\
		    6 & TRE & 17.98 & 356.19 & -35.47 & 1.284 & 17.50 & 8.77 \\
		    7 & TRE & 17.98 & 005.70 & -39.44 & 1.290 & 17.70 & 7.29\\
		    8 & TRE & 17.98 & 016.93 & -43.67 & 1.296 & 18.49 & 3.54\\
		    9 & TRE & 17.98 & 349.71 & -28.01 & 1.301 & 16.54 & 21.36\\
		
		    10 &TCA & 3.24 & 132.80 & +47.98 & 1.304 & 16.78 & 17.18\\
		    11 &TCA & 3.24 & 130.86 & +46.04 & 1.307 & 17.01 & 13.99\\
		    12 &TCA & 3.24 & 136.38 & +52.02 & 1.326 & 16.70 & 18.86\\
		    13 &TCA & 3.24 & 134.20 & +49.89 & 1.346 & 17.28 & 11.19\\
		
		    14 &TCH & 3.24 & 353.66 & -33.89 & 1.617 & 18.78 & 3.39\\
		    15 &TCH & 3.24 & 357.32 & -36.05 & 1.621 & 17.34 & 12.74\\
		    16 &TCH & 3.24 & 002.04 & -38.37 & 1.631 & 19.10 & 2.54\\
		    17 &TCH & 3.24 & 004.45 & -39.33 & 1.633 & 16.80 & 21.28\\
		    18 &TCH & 3.24 & 359.62 & -37.32 & 1.667 & 18.83 & 3.34\\
		\hline
		    \multicolumn{8}{c}{GW170814}\\
		\hline
		    1 & TRE & 17.98 & 034.83 & -52.29& 0.399 & 16.60 & 1.86\\
		    2 & TRE & 17.98 & 046.61 & -35.47& 0.499 & 14.83 & 11.85\\
		    3 & TRE & 17.98 & 041.46 & -35.48 & 0.503 & 16.95 & 1.70\\
		    4 & TRE & 17.98 & 044.01 & -43.84& 0.511 & 18.00 & 0.66\\
		    5 & TRE & 17.98 & 038.36 & -43.89 & 0.516 & 15.93 & 4.43\\
		    6 & TRE & 17.98 & 042.96 & -48.08 & 0.521 & 15.41 & 7.22\\
		    7 & TRE & 17.98 & 036.68 & -48.09 & 0.525 & 17.11 & 1.52\\
		    8 & TRE & 17.98 & 045.38 & -39.68 & 0.578 & 15.58 & 6.91\\
		    9 & TRE & 17.98 & 039.92 & -39.68 & 0.583 & 16.07 & 4.42\\
		    10 & TRE & 17.98 & 041.65 & -52.29& 0.587 & 16.29 & 3.63\\
		    11 & TCA & 3.24 & 034.14 & +48.15 & 3.373 & 16.16 & 23.53\\
		    12 & TCA & 3.24 & 033.10 & +44.43 & 3.378 & 16.14 & 24.00\\
		    13 & TCA & 3.24 & 036.93 & +48.15 & 3.384 & 16.94 & 11.49\\
		\hline
		\end{tabular}
	    \label{table:tab1}
	\end{table*}
\section{IMAGE ANALYSIS AND RESULTS}\label{sec:analysis_results}
\subsection{The Search for Transient Sources}
    We provide here some details on the basic steps described section 2. 
	\begin{enumerate}
	    \item As shown in Figure \ref{fig:Fig2}a and \ref{fig:Fig2}b distortion is still present after the first step. We used the Tycho-2 and 2MASS catalogs to map the shape of the bright stars in  the image. All TAROT telescopes were considered as having a large FoV and we were able to use the same configuration and index files for both Tycho-2 and 2MASS, with skymark diameters ranging from $22'$ to $60'$, to correct for image distortion. We obtained the median of angular separation close to the TAROT pixel scale as shown in Figure \ref{fig:Fig2}c and \ref{fig:Fig2}d.
	
	    \item We used SExtractor with appropriate configurations for each telescope and sensor combination, and with similar output parameters. The extraction threshold was set at $3\sigma$ and the minimum number of pixels above threshold was set to 5. The value of the seeing was obtained directly from the FITS header (it is a part of the initial calibration procedure of TAROT); we used it to discriminate star-like objects from extended sources.

        \item The transient search algorithm was applied to each image. It resulted in lists where sources were classified as unknown sources, particles, or known sources: asteroids, artificial objects or stars in other catalogs.
	
	    \item \textbf{The} possible candidates that were stars and galaxies were automatically rejected by comparing with standard catalogs by a search radius of $10~arcsec$. For every candidate we looked for a possible association in a catalog from the VizieR database. We eliminated asteroids and artificial objects by human vetting, if not already performed automatically during the catalog comparison.
		
		\item For each candidate, photometry was performed and a light curve was computed. None of the candidates appeared to be a credible optical counterpart of the GW event among the 13 objects reported in Table \ref{table:tab2}.
		
    \end{enumerate}
    
	\subsection{Candidate Identification}
	    After the transient search algorithm we had to filter the candidates among artificial objects or cosmic-ray hits.
	\begin{itemize}
		\item Prompt particle events distributed across an image were rejected as they failed to meet the star-like point spread function criterion of SExtractor with appropriate parameters. The fake sources that were still present in the lists of objects were rejected later on the basis of the photometric analysis.
		\item It was possible that some candidates were accidentally eliminated if they were closer to a known source than $10~arcsec$. This prompted for human screening at the end of the whole process.
		\item The algorithm ran poorly with dense star regions such as the center of Milky Way or clusters. The failure of the algorithm came from the detection of faint sources which have magnitude above 18 spread in the background considering the pixel scale and FoV of telescopes. For TCA and TCH the process failed when the number of sources exceeds 15,000 and TRE is around 65,000. In that case it is not possible to run the algorithm, and the processing stops.
		\item Highly distorted images were ignored by the algorithm.
	\end{itemize}

In total 13 candidates where provided by this procedure. They are listed in Table~\ref{table:tab2}. None of these objects appeared to have any connection with a GRB or the GW event. All candidates were rejected by Equation~\ref{eq:3}.

\section{DISCUSSION}\label{sec:discussion}
\begin{figure}[H]
	    \begin{center}
	    \includegraphics[width=1\linewidth]{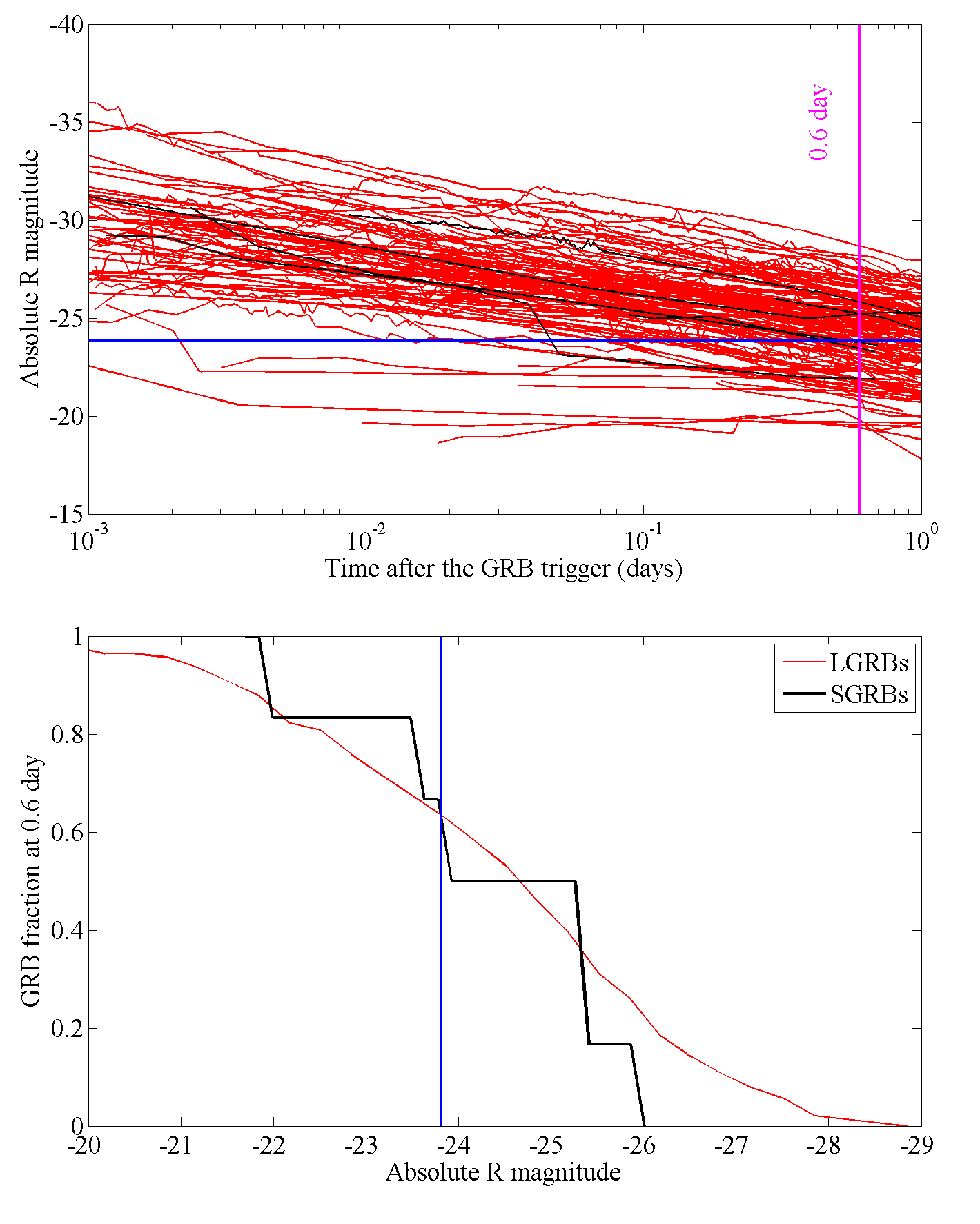}
	    \caption{Upper panel: Optical light curves of 141 long GRBs (red) and of 6 short GRBs (black) observed at 0.6 day after the start of the event, and with known redshifts. The TAROT observation time is 0.6 day (pink line) and the TAROT limiting magnitude is the blue line. Bottom panel: Cumulated fraction of GRBs at 0.6 day derived from the data used for the upper plot.} 
	    \label{fig:gw170814_grbs}
	    \end{center}
    \end{figure}

	    None of the three GW events exhibits an optical counterpart in TAROT images. The best constraint is a limiting magnitude of R=15.0 at 0.6 day after the GW170814 coalescence (conservative values), for which TAROT observed almost the entire 90\% probability contour area. At the distance of the GW event (D$_{GW}$ = 580 Mpc) ~\citep{2018arXiv181112907T} the absolute limiting magnitude from TAROT is M$_{\rm R}$ = -23.8. We collected 141 optical light curves of LGRBs and 6 light curves of SGRBs for which the redshifts are known and we converted them in absolute R magnitude (Figure \ref{fig:gw170814_grbs}). 65\% of them are brighter than M$_{\rm R}$ = -23.8 at the equivalent time of the TAROT observations for GW170814 (i.e. 0.6 day). As a consequence our observations exclude at 65\% an association of GW170814 with a GRB optical counterpart. These limits do not constrain the possible presence of a kilonova or a supernova event possibly associated with the GW source (albeit this is not expected in the case of the coalescence of two black holes).
	    
	    As already mentioned some theories predict an EM emission from BBH mergers. To constrain these theories we computed the $\alpha$ parameter defined in the Equation \ref{eq:1}. Values are given in the Tables \ref{table:tab1}. Conservative values are $\alpha < 1 \times 10^{-5}$, $\alpha < 2 \times 10^{-5}$ and $\alpha < 1 \times 10^{-5}$ for GW150914, GW170104 and GW170814 respectively.
       	The GW170814 gives the most constraining value of $\alpha$ due mainly to the short delay of the observations after the trigger. Moreover only the error box of GW170814 was fully covered by the TAROT observations.
       	
       	The emission process proposed by 
       	\citet{Zhang:2016rli} depends on the parameter $\hat{q}$, which is the fraction of the characteristic charge of the black holes. Limits on Zhang's $\hat{q}$ parameter requires a theoretical work beyond the scope of this paper. Our analysis allows us to establish an observational limit on the fraction $\alpha$ of the total energy that could be emitted as a GRB-like EM radiation. Any model for BBH emission should be bounded by this observational constraint.

     \textbf{As the EM radiation considered here is likely to be emitted from an ultra-relativistic jet, the question of the inclination of the jet along the line of site is important. Here we do not consider the effect of off-axis observations. We expect that the distribution of inclination angles will be known from the observation of BBH mergers during the present and upcoming observing runs.}

\section{CONCLUSION}\label{sec:conclusion}
	We have developed a procedure to automatically process the images from the TAROT telescope network, and applied it to the observations performed in response to GW triggers for three BBH events. The transient search algorithm has proven to be very efficient in providing a list of credible candidates. None of the candidates provided for GW150914,  GW170104 and GW170814 could be associated with a GW source EM counterpart. TAROT observations gives an unprecedented result because we covered almost 100\% of the GW error box only 0.6 day after GW event, comparable to the work of \citet{Doctor:2018ray} who observed 86\% of the GW skymap 1 day after the GW170814 event. Our limits imply that 65\% of known GRB optical light curves should have been detected if optical GRB counterparts are associated to BBH mergers. This does not completely exclude an association between GRBs and BBH mergers but TAROT data considerably reduce the probability of this hypothesis. At least the optical counterpart, if any, should be fainter than that of typical GRBs.
	During their next runs (O3 and later), Virgo and LIGO will feature higher sensitivity combined with smaller, 100~deg$^2$ or less, error boxes. Moreover alerts will be sent within hours, and even less, after the detection of the GW event. As a consequence, TAROT will be able to react very early, providing tight constraints on the optical emission associated with BBH mergers, as well as BNS (Binary Neutron Stars), and hopefully NSBH (Neutron Star Black Hole coalescence). As an example, if we consider the upper plot of Figure \ref{fig:gw170814_grbs}, more than 90\% of GRB optical light curves are brighter than the TAROT limit one hour after the event. In addition, the observation of a large sample of BBH coalescence will enable to study statistically the luminosity (or the absence of) of the possible EM counterpart as a function of the viewing angle.
	We conclude that in a close future, network of large field of view, yet sensitive, telescopes will be able to provide tight constraints on the EM counterpart of binary black hole coalescence.

\begin{table*}[h]
    \caption{The candidates found by the procedure explained in this work. The $\alpha$ parameter represents the efficiency of conversion of GW into EM radiation (see text)}
	\begin{tabular}{@{}lccccccr@{}}
	\hline
		GW event & Observation delay	& RA $(J2000)$	& Dec $(J2000)$ & Mag    & Limit Mag & $\alpha$	\\
		         & After trigger(day)	&(h:m:s)		& (d:m:s)       & (Rmag) & (Rmag)    & ($10^{-5}$)			\\
	\hline
        GW150914 & 2.836  & $07:24:21.00$ & $-68:59:05.90$ & $15.83\pm0.02$  & 17.17 & 1.29\\ 
	    LV trigger: G184098 & 2.857  & $07:33:06.29$ & $-70:24:57.80$ & $13.98\pm0.01$  & 17.17 & 7.14\\
	    2015-09-14 09:50:45 & 2.877  & $07:33:47.07$ & $-69:31:16.10$ & $14.09\pm0.01$ & 16.70 & 6.49\\
	    $\Delta E =$ 3.1$M_{\odot}$& 2.898  & $07:33:22.89$ & $-70:27:40.10$ & $14.67\pm0.03$  & 16.86 & 3.83\\
	    Luminosity distance = 430 Mpc & 2.989  & $07:33:45.71$ & $-69:33:21.10$ & $12.04\pm0.01$  & 17.08 & 44.6\\
	    & 2.947  & $07:25:16.73$ & $-68:50:05.30$ & $15.35\pm0.02$  & 17.32 & 2.08\\
	    & 9.930  & $07:34:56.33$ & $-70:03:24.70$ & $15.03\pm0.01$  & 16.77 & 9.43\\
	    &&& &&& &\\
			
	\hline
	    GW170104 & 1.261  & $23:52:17.83$ & $-37:52:45.08$ & $16.23\pm0.09$  & 17.67 & 2.77\\
	    LV trigger: G268556 & 2.253  & $23:09:18.14$ & $-26:36:52.50$ & $15.26\pm0.04$  & 16.69 & 7.08\\
	    2017-01-04 10:11:59 & 2.270  & $00:58:21.95$ & $+45:30:20.36$ & $15.84\pm0.04$  & 18.10 & 11.94\\
	    $\Delta E =$ 2.2$M_{\odot}$ &&& &&& &\\
	    Luminosity distance = 960 Mpc &&& &&& &\\
	    &&& &&& &\\
	    
	\hline
	    GW170814& 0.522  & $02:41:02.41$ & $-49:42:41.28$ & $17.27\pm0.04$ & 18.65 & 0.13\\
	    LV trigger: G297595 & 0.522  & $02:40:46.12$ & $-49:25:40.31$ & $16.44\pm0.02$ & 17.29 & 0.28\\
	    2017-08-14 10:30:43 & 1.402  & $02:45:11.30$ & $-49:21:27.00$ & $15.85\pm0.07$ & 17.22 & 1.30\\
	    $\Delta E =$ 2.7$M_{\odot}$ &&& &&& &\\
	    Luminosity distance = 580 Mpc &&& &&& &\\
	    &&& &&& &\\
	\hline
	\end{tabular}
	\label{table:tab2}
\end{table*}


\acknowledgments
The TAROT telescope network has been built thanks to the support of the Centre National de la Recherche Scientifique, Institut National des Sciences de l'Univers (CNRS/INSU), and of the Centre National d'Etudes Spatiale, French Space Agency (CNES). It is partially maintained with the kind support of the Observatoire des Sciences de l'Univers (OSU) Pytheas, CNRS - Aix-Marseille University and of the Observatoire de la Cote d'Azur. KN thanks the support of the Kingdom of Thailand, Royal Thai Government Scholarship. RL acknowledges the support of the CNES. 
This work has made use of data from the European Space Agency (ESA)
mission {\it Gaia} (\url{https://www.cosmos.esa.int/gaia}), processed by
the {\it Gaia} Data Processing and Analysis Consortium (DPAC,
\url{https://www.cosmos.esa.int/web/gaia/dpac/consortium}). Funding
for the DPAC has been provided by national institutions, in particular
the institutions participating in the {\it Gaia} Multilateral Agreement.

\facility{TAROT:TCA,TAROT:TCH,TAROT:TRE,LIGO,EGO:Virgo}
\software{Astrometry.net \citep{2010AJ....139.1782L},
SExtractor \citep{1996A&AS..117..393B},
Astropy \citep{2013A&A...558A..33A},
Aladin \citep{2000A&AS..143...33B}
}
\pagebreak
\bibliography{mybibfile}


\end{document}